# Direct Visualization of Irreducible Ferrielectricity in Crystals

Kai Du[1,#], Lei Guo[2,#], Jin Peng[2,#], Xing Chen[1], Zheng-Nan Zhou[1], Yang Zhang[2], Ting Zheng[2], Yan-Ping Liang[2], Jun-Peng Lu[2], Zhen-Hua Ni[2], Shan-Shan Wang[2], Gustaaf Van Tendeloo[3,4], Ze Zhang[1], Shuai Dong[2]*, He Tian[1]*

[1]Center of Electron Microscopy, State Key Laboratory of Silicon Materials, and School of Materials Science and Engineering, Zhejiang University, Hangzhou, 310027, China

[2]School of Physics, Southeast University, Nanjing 211189, China

[3]Electron Microscopy for Materials Science (EMAT), University of Antwerp, Groenenborgerlaan 171, B-2020 Antwerp, Belgium

[4]Nanostructure Research Centre (NRC), Wuhan University of Technology, Wuhan, 430070, China

[#]These authors contribute equally to the work.

*Correspondence to: Email: hetian@zju.edu.cn; sdong@seu.edu.cn;

**In solids, charge polarity can one-to-one correspond to spin polarity phenomenologically, e.g. ferroelectricity/ferromagnetism, antiferroelectricity/antiferromagnetism, and even dipole-vortex/magnetic-vortex, but ferrielectricity/ferrimagnetism kept telling a disparate story in microscopic level. Since the definition of a charge dipole involves more than one ion, there may be multiple choices for a dipole unit, which makes most ferrielectric orders equivalent to ferroelectric ones, i.e. this ferrielectricity is not necessary to be a real independent branch of polarity. In this work, by using the spherical aberration-corrected scanning transmission electron microscope, we visualize a nontrivial ferrielectric structural evolution in BaFe$_2$Se$_3$, in which the development of two polar sub-lattices is out-of-sync, for which we term it as irreducible ferrielectricity. Such irreducible ferrielectricity leads to a non-monotonic behavior for the temperature-dependent polarization, and even a compensation point in the ordered state. Our finding unambiguously distinguishes ferrielectrics from ferroelectrics in solids.**

## Introduction

Ferrielectricity, the equivalent of ferrimagnetism, can be termed as antiferroelectric order but with a switchable polarization, as sketched in Fig. 1a. The first proposed ferrielectric system was a composition of two antiferroelectric compounds $NaVO_3$ and $NaNbO_3$ [1], but subsequent experiments did not support this claim [2]. Despite a long history, the existence of ferrielectricity in solid crystals remains rare, except in liquid crystals [3]. Recently, some solids were claimed to be ferrielectric. For example, in the layered-structure $CuInP_2S_6$, the positions of Cu and In are always opposite within each unit cell (u.c.), which can be recognized as two uncompensated polar sub-lattices with antiparallel alignment [4,5]. In geometric ferroelectrics, such as $Ca_3Ti_2O_7$ and some perovskite superlattices, the displacements of ions are opposite between layers, which can also been recognized as two uncompensated dipole moments [6,7]. Despite the macroscopic and phenomenological analogy as shown in Fig. 1a, it should be noted that there is a key difference between the ferrimagnetic and ferrielectric systems in the microscopic level. Different from the spin moment which can be defined on a single ion, the definition of a charge dipole involves more than one ion and thus may have multiple choices in ionic crystals. By choosing different ions as a dipole unit, multiple dipole values for a sub-lattice can be obtained [8]. In this sense, the Cu-In pair in $CuInP_2S_6$ or bilayer in $Ca_3Ti_2O_7$ can be treated as a dipole unit, which are indeed entangled simultaneously [4,6]. Therefore, these systems are indistinguishable from ferroelectrics, i.e. 'reducible' ferrielectrics, as qualitatively sketched in Fig. 1a. Indeed, $CuInP_2S_6$ only shows one paraelectric-ferroelectric transition temperature, and in most cases those geometric ferroelectrics are coined as 'ferroelectrics' rather than 'ferrielectrics' [5,9]. Similar situations also exist for other ferrielectrics like $Pb_2MnWO_6$ [10]. Conceptually different to these 'reducible' ferrielectrics, in this article, we provide proof of an 'irreducible' ferrielectricity (see Fig. 1a for its conceptual definition of 'irreducible' ferrielectricity) in $BaFe_2Se_3$. We will show unique ferri-characteristics clearly distinguishable from ferroelectricity.

## Results

### $BaFe_2Se_3$: ferrielectric material at room temperature

$BaFe_2Se_3$ belongs to the iron-based superconductor family [11-13] but predicted to be multiferroic under ambient conditions [14]. As shown in Fig. 1b, a $BaFe_2Se_3$ u.c. contains two iron ladders (labelled as A and B). Long-range block-type antiferromagnetism (block-AFM) appears below the Néel temperature $T_N \sim 240\text{-}256$ K [15-19]. The structural tetramerization due to the block-AFM leads to charge dipoles along the $a$-axis and the alignment of dipoles is almost antiparallel but with a tiny canting angle (~5.78° at room temperature) between ladders A and B (schematic displacements are shown in Fig. 1c) [14]. It should be noted that all other members of the 123-series iron selenides (e.g. $BaFe_2S_3$) also own a similar (quasi-) one-dimensional (1D) ladder structure, but only $BaFe_2Se_3$ has the canting ladder characteristic. According to theory [14], a residual polarization along the $c$-axis ($P_c$) is expected in $BaFe_2Se_3$, as a characteristic of a 'reducible' ferrielectric material.

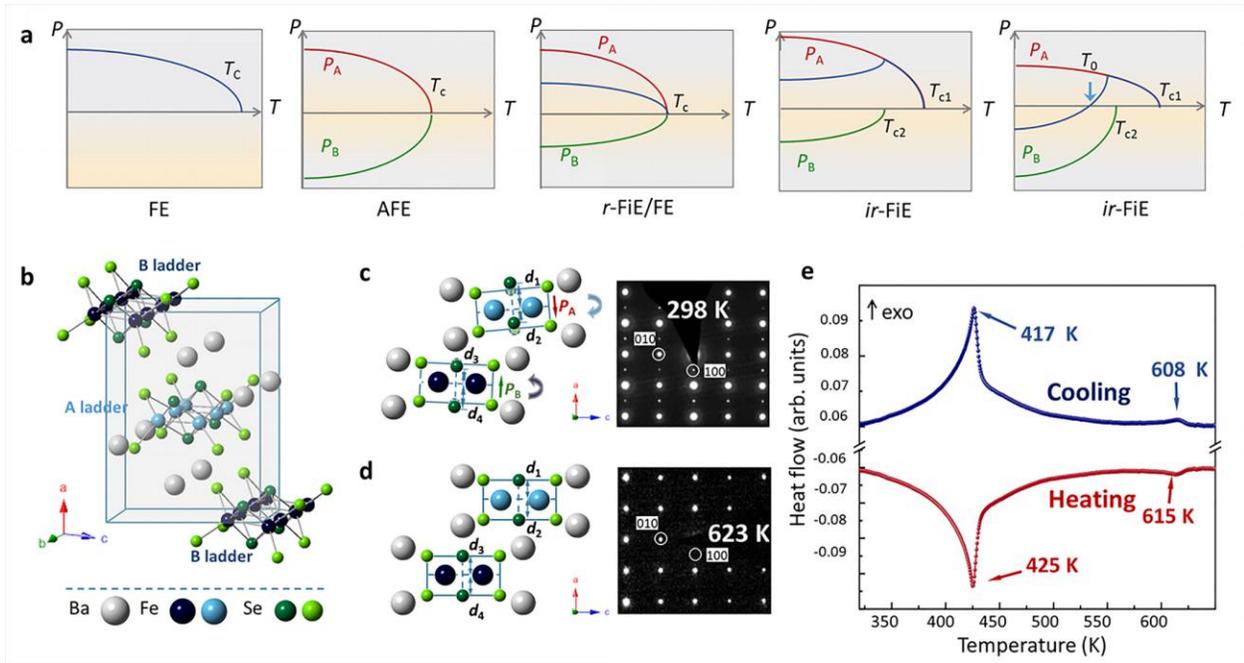

**Fig. 1 Concept of ferrielectricity and candidate material $BaFe_2Se_3$. a**, Polarization ($P$) *vs* temperature ($T$) of ferroelectricity (FE), antiferroelectricity (AFE), reducible-ferrielectricity (*r*-FiE) and irreducible-ferrielectricity (*ir*-FiE), respectively. In *r*-FiE, two polar sub-lattices (A & B) develop synchronously, with one unique critical temperature. This type of ferrielectricity is macroscopically equivalent to ferroelectricity. In *ir*-FiE, the evolution of sub-lattices A & B is out-of-sync, with two critical temperatures (a higher $T_{C1}$ & a lower $T_{C2}$). A compensation point ($T_0$) may appear in some *ir*-FiE systems. Note that any information on the order of the phase transition is not represented in these qualitative cartoons. **b**, Crystal structure of $BaFe_2Se_3$. Each unit is composed of two iron-ladders. **c**, Room temperature structure with a tiny tilting

angle between the ladders A and B. **d**, High temperature structure without tilting. The in-situ selected area electron diffraction patterns are shown. The lattice periodicity along the [100] direction changes from 11.90 Å to 5.96 Å as the temperature rises from 298 K to 623 K. **e,** DSC curves indicate two transitions at ~610 K and ~420 K. The ~610 K transition is a second-order one with a step-like behavior, while the ~420 K transition is a first-order one with a peak.

The tilting of the iron ladders gradually disappears with increasing temperature up to ~600 K, leading to a high symmetric *Bbmm* phase, according to X-ray diffraction [16]. Our in-situ selected area electron diffraction (SAED) results give distinctive patterns (Fig. 1c *vs* Fig. 1d), confirming the disappearance of the ladder canting at high temperature. Differential Scanning Calorimetry (DSC) measurements (Fig. 1e) confirm the phase transition occurring at ~610 K as a second-order one. Besides, there is a first-order phase transition at ~420 K, which was also evidenced in Ref. 16 but its real origin remains a puzzle. In addition, the resistivity behavior also supports this first-order transition (Supplementary Figure 1). Since our neutron powder diffraction (NPD) data confirm that only one block-AFM transition appears at 250 K, obviously, the transition at ~420 K can be excluded as a magnetic-ordering behavior (Supplementary Figure 2a).

Using a spherical aberration-corrected scanning transmission electron microscope (Cs-STEM), we were able to determine the subtle structure evolution of the 1D Fe chains as well as the associated Se's. Then the structure of $BaFe_2Se_3$ can be directly measured atom column by atom column, which reveals an unexpected irreducible ferrielectricity going beyond the theoretical expectations. The STEM image of $BaFe_2Se_3$ along *b*-axis and *c*-axis was shown in Supplementary Figure 3. The STEM results at room temperature are summarized in Fig. 2. Interestingly the structural tetramerization already exists at room temperature and, unexpectedly, the intensities for ladders A and B are not equivalent, creating a strong ladder and a weak ladder within a unit cell. The inequivalent features in strong ladders and weak ladders are embodied in magnitude of displacements, as shown schematically in Fig. 2a and in the high-angle annular dark-field scanning transmission electron microscopy (HAADF-STEM) images along the *b* (Fig. 2b and 2d) and *c* axis (Fig. 2c and 2e). The line profiles along these two ladders are shown in Supplementary Figure 4. A displacement vector-mapping algorithm was implemented on the cross-sectional HAADF-STEM images to measure the local displacement of the atoms. Based on the statistics of about 300 data for each length, the Fe atoms displacement in the strong ladders are stronger than that in the weak ladders (see Supplementary Figure 5 for an example). Considering that the Fe-block tetramerization could induce Se ions displacement along the *a*-axis [14] (as clearly indicated in Fig. 2b and 2d), such inequivalence of Fe displacements will lead to a residual polarization along the *a*-axis ($P_a$), which is larger than the expected $P_c$. The arrow is added to make it more visible. The uncovered zoom-in images (raw data) have been shown in Fig. 2(b-c). Therefore, $BaFe_2Se_3$ is a room temperature ferrielectric with polarization mainly along the *a*-axis, rather than the expected low temperature ferrielectric with polarization along the *c*-axis [14].

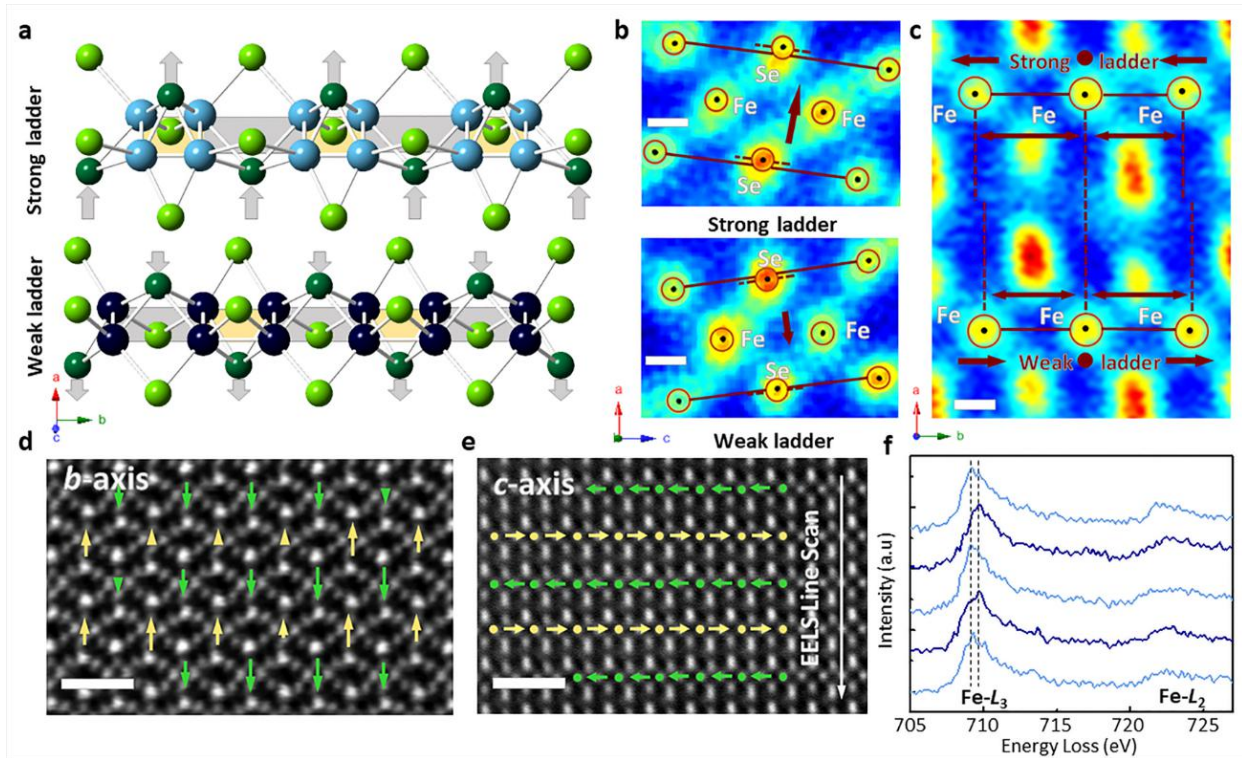

**Fig. 2 Atomic displacements in the ladders. a**, Schematic displacements of Fe and Se ions indicated by arrows and amplified in magnitude. Light blue: strong ladder with larger polar bias of Se ions; Dark blue: weak ladder. **b** and **c**, Color-enhanced HAADF-STEM images of $BaFe_2Se_3$ along *b*-axis and *c*-axis. The strong and weak ladders are distributed in an alternating order. Scale bar, 1 Å. **d** and **e**, Superposition of the HAADF image and the polar map of Se atoms along *b*-axis and Fe atoms along *c*-axis. The yellow and green arrows there represent the atoms displacements in strong and weak ladders respectively. Scale bar, 1 nm. **f**, The EELS line scan is across the white line in (**e**). EELS measurements indicate the charge ordering pattern.

The averaged Fe $L_{2,3}$ edge spectrum from neighboring Fe chains is presented. A Fe-$L_3$ peak shift of approximately 0.4 eV between neighboring Fe chain is observed.

To explore the origin of this unbalanced ladders, monochromated electron energy loss spectra (mono-EELS) were acquired to demonstrate the underlying charge modulation as shown in Fig. 2f. Using the monochromatror we reach an energy resolution of 0.3 eV, which is enough to detect subtle changes in the fine structure of the EELS excitation edges. The averaged Fe-$L_3$ edges in strong chains and in weak chains show significant differences in their ELNES. Comparing with Fe reference spectra [20], the valence states of the strong and the weak chains in BaFe$_2$Se$_3$ are different, and the Fe-$L_3$ peak shift between them is approximately 0.4 eV (shown as the distance between two dashed line in Fig. 2f). Note that the EELS have an ability to reflect a valance change, but the absolute valance identification is still challenging, since the absolute energy position always has several hundred meV uncertainty. Such charge disproportion is not unusual in correlated electron systems, especially in Fe-based oxides and fluorides, e.g. Fe$_3$O$_4$ and LiFe$_2$F$_6$ [21, 22], although it has not been reported in selenides before. Nominally, the valences of Fe can become $+(2+\delta)$ and $+(2-\delta)$ for the two sublattices, and a proper $\delta$ (around 0.15 according to our fitting results) can lead to the structural tetramerization following the idea of Peierls transition [14]. It should be noted that the non-integer valences are possible in iron selenides, e.g. in KFe$_2$Se$_3$ and KFe$_2$Se$_2$ [23-25]. Therefore, charge-ordering, i.e. difference of local electron density, can be the key ingredient for the unbalanced structural tetramerization and affiliated polarization, which needs deeper investigation in future.

**Variation of the BaFe$_2$Se$_3$ structure with various temperature**

Then it is interesting to know its ferrielectric $T_C$. To characterize the structural tetramerization, the difference ($\delta$) of nearest-neighbor Fe-Fe bond length ($d$) along the ladder direction is measured as a function of temperature (Fig. 3a). An in-situ heating experiment was performed using a DENSsolutions SH30 system to be able to measure over a wide temperature range. A lamella of BaFe$_2$Se$_3$ was transferred onto specialized chips using a probe-assistance method (Supplementary Figure 6), then the sample was heated to the set temperature by resistance heating. The displacement vector-mapping algorithm was implemented on the STEM images, and each $\delta$ is obtained by averaging around 300 measurements. The difference of Fe-Fe bond length as well as the tilting angle between the ladders at high temperature vanishes ~600 K, and the in-ladder tetramerization almost drops to zero, implying a high symmetric nonpolar phase (in agreement with the DSC curve shown in Fig. 1e and the X-ray data in Ref. [16]). The discrepancy between strong and weak ladders disappears in this high symmetric phase, as also demonstrated by above SAED result (Fig. 1d). With decreasing temperature (e.g. at 473 K), unexpectedly, tetramerization emerges in one sublattice of ladders but not in the other, which is a unique characteristic of irreducible ferrielectricity, leading to an emergence of $P_a$ (Fig. 3b) [calculated by density functional theory (DFT)] [26]. Meanwhile, the ladders become tilting (Fig. 3c compared with Fig. 3d). With further decreasing temperature (e.g. at 423 K), tetramerization emerges in both ladders with different intensities and slopes, implying that the first-order transition occurring at ~420 K corresponds to the starting of tetramerization of the weak ladders. At ~373 K, the intensities of tetramerization are close to identical between both ladders, resulting in an (almost) canceled $P_a$, i.e. the unique compensation point ($T_0$) of irreducible ferrielectricity. Below $T_0$, the difference between the two ladders increases with decreasing temperature, leading to a reentrance of $P_a$. It is worth to emphasize that the direction of polarization (i.e. the roles of strong and weak ladders) may be probably reversed across $T_0$ according the tendency (see the last panel of Fig. 1a), although our variable temperature measurements can not continuously track the evolution of individual ladder over so large temperature range. If so, the strong ladder A (weak ladder B) becomes the weak ladder A (strong ladder B) when temperature crossing $T_0$. More color-enhanced HAADF-STEM images acquired at different temperatures along the *c*-axis are shown in Supplementary Figure 7. Statistics of the deviation of Fe-Fe bond length in the strong and weak chain at different temperatures could be seen in Supplementary Table 1.

Since the ferrielectric polarization can not be directly measured using electrical methods (and even pizeoelectric force microscopy) in the current stage due to the high leakage of the samples (considering the very small band gap ~0.13-0.178 eV [17, 18]), an optical second harmonic generation (SHG) experiment is employed to characterize the polarity of the materials (see Methods for experimental details). Our SHG signal (Fig. 3b) demonstrates its polarity below ~600 K. Most importantly, the non-monotonic evolution of the SHG signal unambiguously matches the DFT calculated polarization, including the possible compensation point at ~380-400 K, which provides a very strong evidence to support our STEM data.

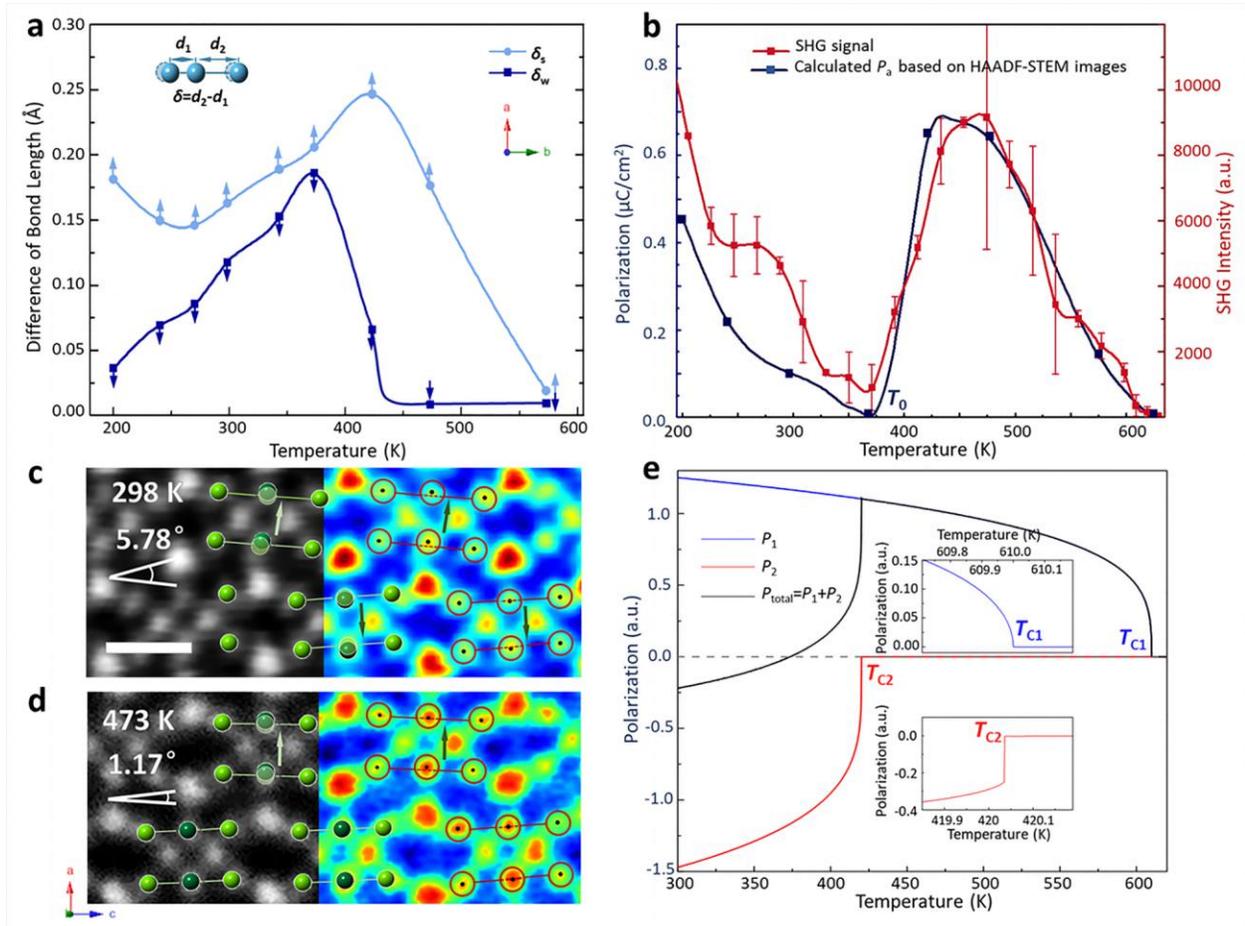

**Fig. 3 Temperature-dependent evolution of BaFe$_2$Se$_3$. a**, The measured structural tetramerization. The light blue and dark blue lines correspond to the strong and weak ladders respectively. The affiliated arrows denote the corresponding dipoles. **b**, Left axis: the polarization along the *a*-axis (DFT calculated using the STEM structural data). Right axis: the SHG signal, which mainly reflects the evolution of $P_a$. The overall evolutions of the polarization and SHG signal qualitatively match. More SHG data and explanation can be found in Supplementary Figure 8 and Figure 10. **c** and **d**, Color-enhanced HAADF-STEM images acquired at 298 K and 473 K along the *b*-axis. The Se atoms displacement in the strong chain is larger than that in the weak chain at 298 K; this is consistent with the difference of the Fe chain evolution magnitude at 298 K. Also, the tilt angle between ladders varies from 5.78° to 1.17° at high temperature. Scale bar, 5 Å. **e**, Simulated evolution of order parameters with a most simplified Landau-type free energy formula. Inserts: the magnified views near the phase transitions, which indicate a second-order transition at $T_{C1}$ and a first-order transition at $T_{C2}$.

The irreducible ferrielectricity of $BaFe_2Se_3$ can be qualitatively described by Landau theory with a coupling between two ladders ($P_A$ & $P_B$). The most simplified free energy formula can be written as:

$$F=\alpha_A(T-T_A)P_A^2+\beta_A P_A^4+\gamma_A P_A^6+\alpha_B(T-T_B)P_B^2+\beta_B P_B^4+\gamma_B P_B^6+\alpha_{AB}P_A \cdot P_B, \qquad (1)$$

where the first to six items are the standard Laudau-Ginzburg-Devonshire type energy expression up to the sixth power for sub-lattices A and B, while the last item is the antiferroelectric coupling between two sub-lattices. All coefficients except $\beta_2$ are positive and the small canting angle between $P_A$ and $P_B$ is neglected. Without fine tuning of the coefficients (see Methods for details), the simulated evolution of the polarization (Fig. 3e) is qualitatively reproducing the non-monotonic experimental behavior, implying the correct main physics captured in the model. Consistent with the DSC data, the high temperature transition is a second-order one, while the low temperature transition is a first-order one due to the negative $\beta_2$.

**Manipulation of $BaFe_2Se_3$ by an external electric field**

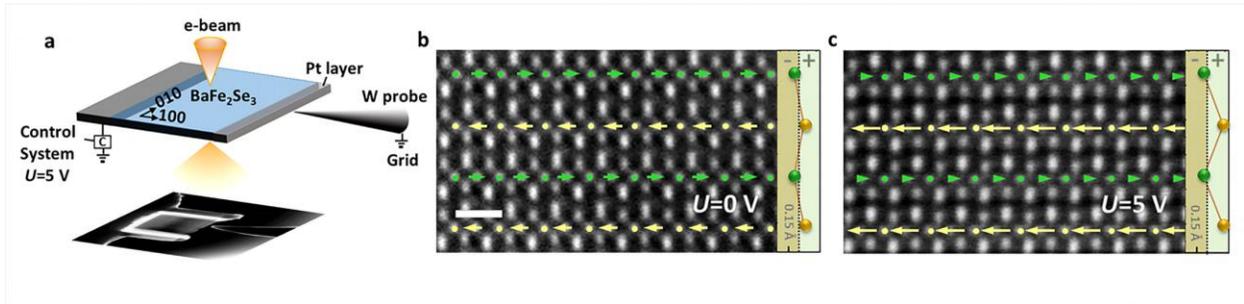

**Fig. 4 Manipulation of the ferrielectric polarization in $BaFe_2Se_3$. a**, Schematic diagram of the electric field experiment. The in-situ biasing experiment was performed with an input voltage $V_{DC}$=5 V. To capture the structural changes, a real-time analysis is performed in STEM mode. **b** and **c**, Comparison of the local dipoles with/without applied voltage. The arrows denote the displacements of 50% of the Fe ions, while the rest of the Fe ions are treated as reference points. The difference between the strong/weak ladders is further enhanced by the electric field. Arrows are added to make the displacements more visible. The uncovered zoom-in image (raw data) can be found in Fig. S9. Scale bar, 5 Å.

As a polar material, electric field tuning of the polarization is a fundamental function. However a direct measurement of the electric hysteresis loop is technically challenging in the current stage due to serious leakage, but we have successfully developed an inspiring in-situ technique to manipulate the polar structure at an atomic scale by applying an electric field. An electrical bias was applied between a tungsten tip, which acts as a mobile electrode, and the lamella of $BaFe_2Se_3$, which is connected to ground (as schematically shown in Fig. 4a). To capture the changing of atoms positions, the real-time crystal structure with atomic-scale spatial resolution was characterized in STEM mode; the access to dynamic structural information provide a clear picture of the evolution under an external electric field. As shown in Fig. 4b-c and Supplementary Figure 9, the tetramerization is significantly enhanced by applying an electric field along the $a$-axis. This implies a significant enhancement of polarization under the electric field. However, the reversal of $P_a$ has not been achieved, implying a large coercive field.

Previous experiments reported the space group *Pnma* for $BaFe_2Se_3$ at room temperature [15-17], which is nonpolar and does not allow the tetramerization. A recent work reported the space group *Pmn2$_1$* (a subgroup of *Pnma*) at room temperature, allowing the tetramerization and polarization [27]. Another recent X-ray diffraction work reported the space group *Pmn2$_1$* at 300 K but *Pm* (a subgroup of *Pmn2$_1$* allowing the inequivalent ladders) at 150 K [28]. In fact, the patterns of NPD (or X-ray) are very subtle among these space groups (see Supplementary Figure. 10 and Supplementary Figure11), which thus can not distinguish these structures precisely. Instead, our STEM technique is more suitable to monitor these subtle distortions of inner coordinates. According to our studies, below $T_{C1}$ the accurate space group should already be *Pm*, allowing the asymmetry between two ladders. The transition from *Bbmm* to *Pm* at $T_{C1}$ is a second-order one while the first-order transition at $T_{C2}$ does not change the symmetry. Other high-resolution technique, such as synchrotron X-ray diffraction, may be helpful to verify our STEM results in future.

**Discussion**

Finally, it should be noted that the irreducible ferrielectrics is not limited to $BaFe_2Se_3$ but with broader interests. For example, as an important branch of multiferroics, $TbMn_2O_5$ and other 125-type manganites showed strange 'ferroelectric' behavior of polarization as a function of temperature or magnetic field [29,30], including the compensation point of polarization. The real mechanism is that the ferroelectric contributions in TbMn$_2$O$_5$ are from three out-of-sync sources according to the SHG measurement [30].

Even though, our current work on BaFe$_2$Se$_3$ is not a marginal extension of TbMn$_2$O$_5$. The polarity in TbMn$_2$O$_5$ is magnetism-driven, i.e. it is a so-called type-II multiferroic material, instead of a proper ferroelectric material. It is not rare for a magnetic system to have sequential magnetic phase transitions. In this sense, the nontrivial evolution of polarization in TbMn$_2$O$_5$ is just a secondary effect of magnetic evolution, which occurs at very low temperature (<40 K) and gives a very weak signal of polarization (~<0.04 μC/cm$^2$ and ~<0.15 μC/cm$^2$) [29, 30]. In our case, the ferrielectricity is not magnetism-driven but a primary polar property, which occurs above room-temperature (15 times of TbMn$_2$O$_5$) and with a much larger polarization (5-15 times of TbMn$_2$O$_5$). In addition, limited by its very weak polarization signal, the experimental measurements of TbMn$_2$O$_5$ can only rely on the pyroelectric method, which can lead to a net polarization but the microscopic facts of different contributions were mostly by suspecting or indirect derivation from SHG signals. Instead, our current work, powered by the advanced in-situ STEM techniques and thanks to the strong signal of BaFe$_2$Se$_3$, the microscopic evolution of two contributions can be visualized directly, leading to a more decisive conclusion. In fact, although the nontrivial polarization of TbMn$_2$O$_5$ has been known for decades, it is more likely to be recognized as a type-II multiferroics with strange ferroelectric behavior. Our work will lead to a re-look at the irreducible ferrielectricity, including that in TbMn$_2$O$_5$.

The irreducible ferrielectricity combines both characteristics of ferroelectricity and antiferroelectricity, making these systems having more degrees of freedom to be controlled. For example, by tuning the amplitudes of sub-lattice polarizations near the compensation point, the macroscopic polarization can be easily switched, without the reversal process of dipole moments as required in ferroelectric cases. Moreover, complex ferroelectric+antiferroelectric domain structures may be expected in ferrielectrics [31], which deserve further studies.

In summary, employing spherical aberration-corrected STEM with sub-angstrom resolution, the structural evolution of BaFe$_2$Se$_3$ has been characterized in detail. Highly interesting phenomena, beyond previous experimental observations and theoretical predictions, have been detected and analyzed. First, BaFe$_2$Se$_3$ is a room temperature polar material. Second, combined with EELS analysis, the origin of its structural tetramerization is demonstrated to be driven by the local electron density, not the previously expected block-type antiferromagnetism. Third, most importantly, the evolution of the two ladders in BaFe$_2$Se$_3$ does not behave synchronously, leading to irreducible ferrielectricity. The compensation point, a unique fingerprint of irreducible ferrielectricity, is observed. The irreducible ferrielectricity reported here is conceptually different from previously reported reducible ferrielectricity which is actually equal to ferroelectricity. The irreducible ferrielectricity in BaFe$_2$Se$_3$ acts as the primary effect, leading to a stronger impact to the community to re-investigate this independent branch of polarity. More functionalities are promisingly expected in future based on irreducible ferrielectricity, e.g. the magnetic-field-tunable polarization as demonstrated in TbMn$_2$O$_5$.

## Methods

**Material synthesis.** High-quality BaFe$_2$Se$_3$ single crystals were grown by the self-flux technique starting from an intimate mixture of Ba pieces, Fe granules, and Se powders with an atomic ratio of 1: 2 : 3. Then the starting materials were put in a carbon crucible and sealed in the quartz tube with partial pressure of argon. The quartz tube was first heated to 420°C at a rate of 1 °C/min, held for 12 h, and then annealed at 1150 °C for another 24 h. After that, the quartz tube was slowly cooled down to 750 °C at a rate of 3°C/h. Finally, the quartz tube was cooled down to room temperature naturally and the strip-like BaFe$_2$Se$_3$ single crystals with a typical size of 3.0× 1.0 × 0.5 mm$^3$ and shiny surfaces can be obtained.

**Macroscopic properties measurements.** Differential scanning calorimetry (DSC) experiments were performed with Maia DSC 200 F3. Measurements were performed on heating and cooling with a rate of 10 K·min$^{-1}$. The sample is encapsulated in a standard Al crucible using argon stream as the protecting gas. XRD measurements were performed on Rigaku Smartlab3 with Cu $K_\alpha$ radiation. In the SHG measurements, the incident laser with a wave length of 800 nm is perpendicular to the cleavage (100) plane and the reflected light at 400 nm is collected. The polarization of the incident laser is controlled by a half wavelength plate. Then the canting polarization along the *c*-axis can be monitored, which is in proportional to the main component of polarization along the *a*-axis. The SHG signal in Fig. 3b was measured with the polarization of the incident laser (i.e. the electric field component *E*) in the *bc* cleavage surface (Fig. S10c) of the BaFe$_2$Se$_3$ crystal, while the angular-dependent results with rotating *E* can be found in Fig. S8. Magnetic measurements were carried out in a vibrating sample magnetometer (VSM) integrated in a Physical Property Measurement System (PPMS-9, Quantum Design) up to 600 K. Neutron Powder Diffraction (NPD) data were collected on a High-intensity Powder Diffractometer Wombat at Australian Nuclear Science and Technology Organization (ANSTO) with λ=2.41 Å, between 10 K and 500 K. Resistivity was measured using Keithley 4200A-5CS.

**Conventional and scanning transmission electron microscopy.** Samples was cut into lamellas by Focused Ion Beam (FEI Quanta 3D FEG) for the observation of electron microscope. We use spherical aberration correction electron microscopy (FEI Titan G2 80-200 ChemiSTEM, 30 mrad convergence angle, 0.8 Å spatial resolution) to acquire atomic resolution HAADF-STEM images of $BaFe_2Se_3$'s cross section from different directions and the image noise was corrected using Digital Micrograph. All STEM images in this work are filtered in Fourier space using a grid mask to select for the lattice frequencies and by low and high pass annular filters to remove the zero frequency and high frequency noise above the information transfer limit. Electron Energy Loss Spectroscopy (EELS) test was also performed on $BaFe_2Se_3$ to verify whether there are changes in the valence state of iron. SAED patterns (selected area electron diffraction) obtained on FEI Tecnai G2 F20 S-TWIN are used to verify the analysis on the local evolution in the statistical sense. Some additional details should be mentioned: 1) As to exclude the influence of microscope artifacts, some steps have been taken. To minimize the influence of sample drift, most of the microscopy data for quantitative analysis are acquired under the condition of drift below 1 Å $min^{-1}$. 2) The STEM image were acquired from the mutually perpendicular directions, and the analysis results of bond length show no obvious difference between them. Thus, the possibility of STEM scanning direction as the main origination of the observed phenomenon can be excluded. 3) High frequency vibration of imaging would be another potential influence factor. Therefore, a technique of ultra-fast acquirement was employed. Tens of images were quickly acquired in the same local region, and most HAADF images shown in this article are overlaid based on such image series, the effect of the specimen drift and beam vibration were significantly reduced and the signal-to-noise ratio of the HAADF images was improved, simultaneously. (4) Aiming at avoiding oxidation, experiments were performed as soon as $BaFe_2Se_3$ was taken out of glove box which offer protecting gas. (5) In consideration of that potential slight damage caused by ion beam in FIB, we minished parameters including voltage and electric current of ion beam down to 2 kV/27 pA to minimize the negative and unnecessary surface damage. For $BaFe_2Se_3$, the antiferromagnetic ordering temperature is 250 K (Supplementary Figure 2), while most of STEM data are measured far above this temperature. Thus, the magnetic fields (from Cs-STEM) effect to polar distortion is negligible in the high temperature paramagnetic region.

**In-situ study.** The in-situ heating experiment was done on DENS solutions SH30 system in order to carry out the experiment in a wide temperature range. The Nano-Chip we used could control the temperature environment locally on the device via the 4-point-probe. Its highest temperature accuracy and stability is 0.001 °C. The experimental data at low temperature was obtained by a demo low-temperature sample holder made by DENS. The in-situ biasing was done on Hysitron PI-95 TEM PicoIndenter, the input voltage $V_{DC}$=5 V was applied between the sharp conductive tip and the sample using a function generator. To capture the changing of atoms positions, the real-time crystal structure is characterized in STEM. The atomic-scale spatial resolution of STEM and the access to structural information provide a clear picture of the evolution under external electric field.

**Specimen transfer onto biasing chips.** To measure the performance of $BaFe_2Se_3$, as shown in Supplementary Figure 6. After conventional FIB process of welding the lamella onto needle and transfer it near the surface of chips, additional confined Pt pad was deposited to contact the lamella onto the chip surface. The lamella thickness was to about 2 μm. Low current down to 10 pA was used to polish the surface of chips after detaching the needle to reduce the amount of redeposited material resulting from the previous contacting process.

**DFT calculation.** The DFT calculation was performed based on the projector augmented-wave (PAW) potentials and Perdew-Burke-Ernzerhof exchange function as implemented in Vienna *ab initio* simulation package (VASP) [32-34]. The plane-wave energy was 500 eV. The experimental structures at different temperatures were used and the Cx-type antiferromagnetism is adopted for simplify (since here the polarization is not driven by magnetism). Brillouin zone integration was obtained using a 6×3×4 *k*-point mesh. The standard Berry phase method is adopted to estimate the ferroelectric polarization [35], while the intuitive point-charge-model provides similar results.

**Laudau-Ginzburg-Devonshire model.** To fit the experimental phase transitions, $T_A$=610 K & $T_B$=420 K, are used. Noting $T_{C1}$=$T_A$ for the second order phase transition, and $T_{C2}$ is close but slightly higher than $T_B$ for the first order transition. To simulate Fig. 3e, $\alpha_1$=1 as the unit, and $\beta_1$=5, $\alpha_2$=4.5, $\beta_2$=-5, $\gamma_1$=$\gamma_2$=40. The negative $\beta_2$ is essential for the first-order transition around $T_{C2}$ and the differences between $\alpha_1/\alpha_2$, $\beta_1/\beta_2$ originate from the charge disproportion. $\alpha_{12}$ can be a small quantity, e.g. 0.001.

**Determining the position of atoms.** Polarization mapping here was performed by calculating ion displacements in HAADF-STEM images. On account of the fact that the bright area of every atom in HAADF image is actually too large for us to determine where the center of atom is, a mathematical method involving Gaussian Fitting based on Matlab is essential to ascertain the accurate position of every atom. Gaussian Fitting could give an accurate position of atom according to the brightness of every atom.

Determining atomic positions by fitting each atom site using a spherical Gaussian algorithm in Matlab is a common method. For double-check of our conclusions, we used two different softwares: CalAtom [36, 37] and StatSTEM [38] and compare their calculating results. The average value distribution of data using these two softwares are keeping an exact consistency. For example, for the Fe-Fe bond length of strong ladder at room temperature, CalAtom reveals that they are 2.71 Å and 2.55 Å; the StatSTEM reveals that they are 2.72 Å and 2.54 Å Besides, each bond length acquired from atoms-position-determination software are based on the statistics value of about 300 data, thus they are statistically meaningful. Furthermore, multiple images were recorded for most of the data we exhibit. The multiple images were averaged in order to reduce noise and artifacts induced by possible random drifts of the sample. In the average procedure, the aliment of these images was done by minimizing the shift of an individual image relative to the averaged image using an iterative rigid alignment method. Another methods to measure the artifacts in STEM images is to determine the displacements of atoms in which no off-center displacements would happen [39]. On the basis of this method, Lu, L. *et al.* measured the STO layers [39] with rms=4.8 pm, we observed the STO layers with rms=3.2 pm using CalAtom software.

**Data availability.** The datasets generated during the current study are available from the corresponding author on reasonable request.

**Acknowledgments** We acknowledge the National Natural Science Foundation of China (Grant Nos. 11834002, 11674055, and 11234011), National Key R&D Program of China 2017YFB0703100, and the 111 Project (Grant No. B16042). K.D. acknowledges the China Scholarship Council (CSC, No.201806320230) for sponsorship and 2019 Zhejiang University Academic Award for Outstanding Doctoral Candidates.. We thank Prof. Fang Lin for providing guidance on calculating atoms position and Dr. Andrew Studer for performing neutron powder diffraction. We thank Prof. Sang-Wook Cheong, Prof. Zhigao Sheng, Prof. Qianghua Wang, Prof. Meng Wang, Prof. Renkui Zheng, Prof. Takuya Aoyama, Dr. Zhibo Yan, and Dr. Meifeng

Supplementary Information for

# Direct Visualization of Irreducible Ferrielectricity in Crystals

Kai Du[1#], Lei Guo[2#], Jin Peng[2#], Xing Chen[1], Zheng-Nan Zhou[1], Yang Zhang[2], Ting Zheng[2], Yan-Ping Liang[2], Jun-Peng Lu[2], Zhen-Hua Ni[2], Shan-Shan Wang[2], Gustaaf Van Tendeloo[3,4], Ze Zhang[1], Shuai Dong[2]*, He Tian[1]*

[1]Center of Electron Microscopy, State Key Laboratory of Silicon Materials, and School of Materials Science and Engineering, Zhejiang University, Hangzhou, 310027, China

[2]School of Physics, Southeast University, Nanjing 211189, China

[3]Electron Microscopy for Materials Science (EMAT), University of Antwerp, Groenenborgerlaan 171, B-2020 Antwerp, Belgium

[4]Nanostructure Research Centre (NRC), Wuhan University of Technology, Wuhan, 430070, China

[#]These authors contributed equally.

*Correspondence to: Email: hetian@zju.edu.cn; sdong@seu.edu.cn;

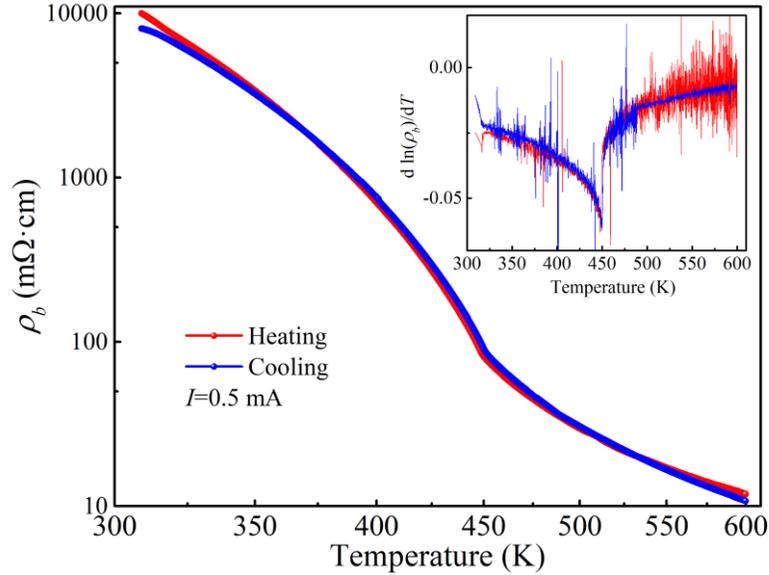

**Supplementary Figure 1** Temperature dependence of the resistivity ($\rho_b$). Insert: the derivative of $\ln(\rho_b)$. The resistivity changes its behavior near 450 K, which also supports the scenario of a first-order transition. The measuring current is along the b-axis. Noting this material is highly conductive (although it is not a metal) with a very small experimental band gap (e.g. 0.178 eV or 0.13 eV) [17, 18], it is technically challenging to directly measure its small polarization (~0.1 μC/cm$^2$) hysteresis loop, i.e. to demonstrate the switching of polarization. Noting that the anomaly of resistivity occurs at a little higher temperature than $T_{C2}$ of DSC (~417-425 K) for the following possible reasons. First, usually thermometers for different experiments are not well relatively-calibrated, especially for the high-temperature range. Second, maybe for such a first order transition, resistivity always changes at a little higher temperature, according to the SHG result of Ref. [27].

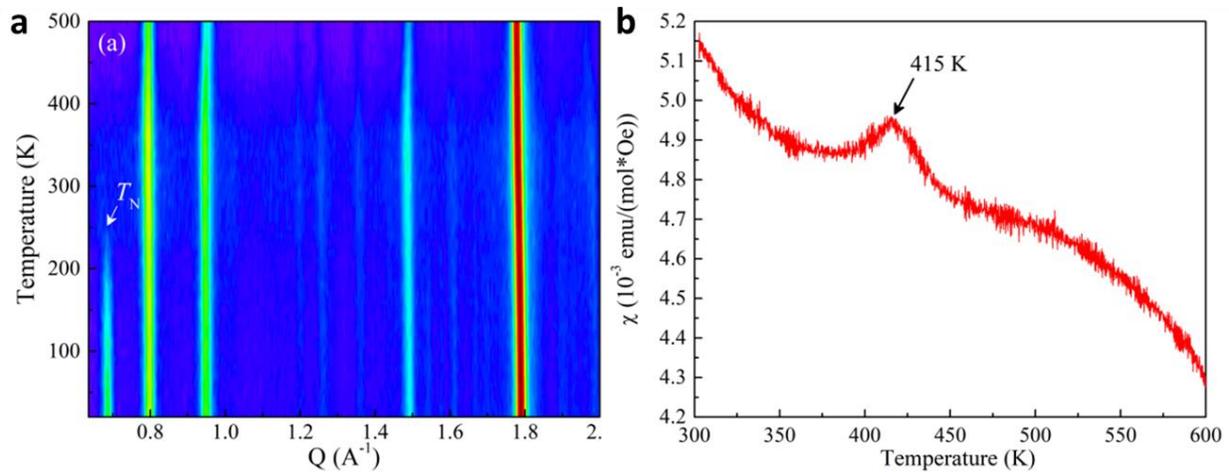

**Supplementary Figure 2** Macroscopic physical properties. (a) Contour plot of the neutron powder diffraction spectrum, collected on the high-intensity powder diffractometer WOMBAT at the Australian Nuclear Science and Technology Organisation (ANSTO) with λ= 2.41 Å, between 10 K and 500 K. The magnetic peak at $Q=0.7$ Å$^{-1}$ for block-type antiferromagnetism appears below ~250 K, in agreement with literature. No other long-range magnetic order is observed in the whole spectrum from $Q=0.63$-$4.75$ Å$^{-1}$. (b) Magnetic susceptibility peaked at ~415 K, measured under 1 T field. Neutron powder diffraction confirms the appearance of long range magnetic ordering below ~250 K, in agreement with previous report of (~255 K) [17, 18] and (~256 K) [15]. Thus, the peak of the magnetic susceptibility at ~415 K (appearing only under strong magnetic field, e.g. ~1 T) should be a side effect of a polar structural transition, instead of a magnetic phase transition.

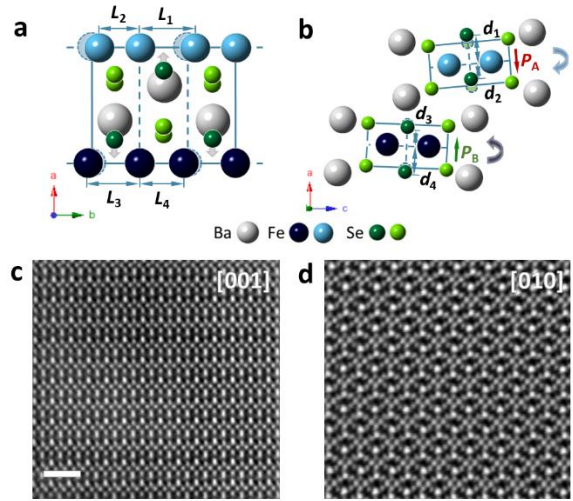

**Supplementary Figure 3** STEM images and sketches of BaFe$_2$Se$_3$ along (a) and (c) the [001] direction, (b) and (d) [010] direction. BaFe$_2$Se$_3$, having an orthorhombic crystal structure, was analyzed with different temperature and external bias. Samples were cut into lamellas with the widest faces perpendicular to the crystal principal axis ([010] and [001]) by Focused Ion Beam for the observation in the electron microscope. Since the BaFe$_2$Se$_3$ is easily dissociated along the [100] plane which could be observed easily, it is convenient to judge orientation and perform further analysis of the accurate crystal structure on the basis of HAADF images along a target direction. Scale bar, 1 nm.

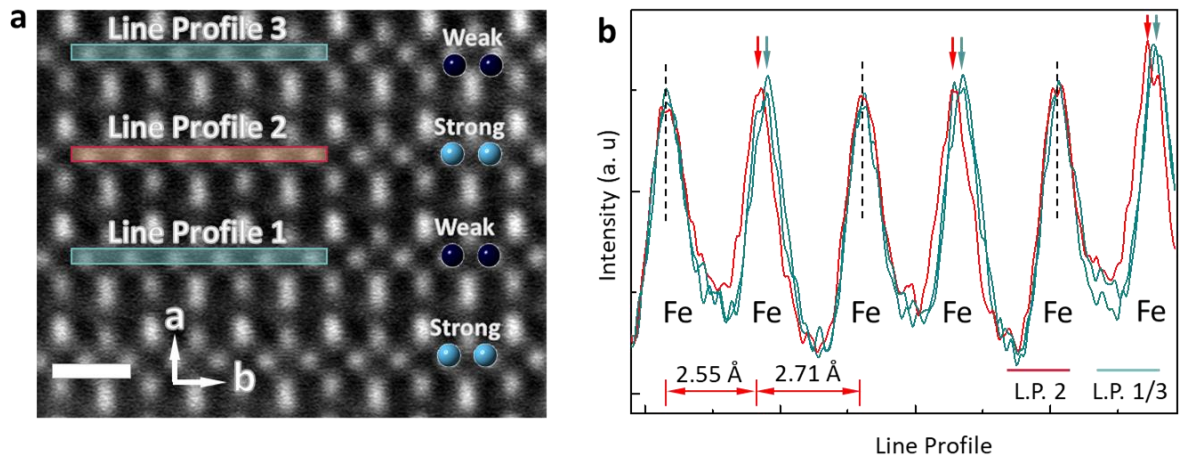

**Supplementary Figure 4** STEM image of BaFe$_2$Se$_3$ along the *c*-axis and corresponding line profiles. (a) HAADF-STEM images along the *c*-axis. (b) Corresponding line profiles which are extracted from the red and cyan boxes in (a), which represent weak and strong ladders respectively. The difference between the bond length of strong and weak ladders is clearly observed. Scale bar, 5 Å.

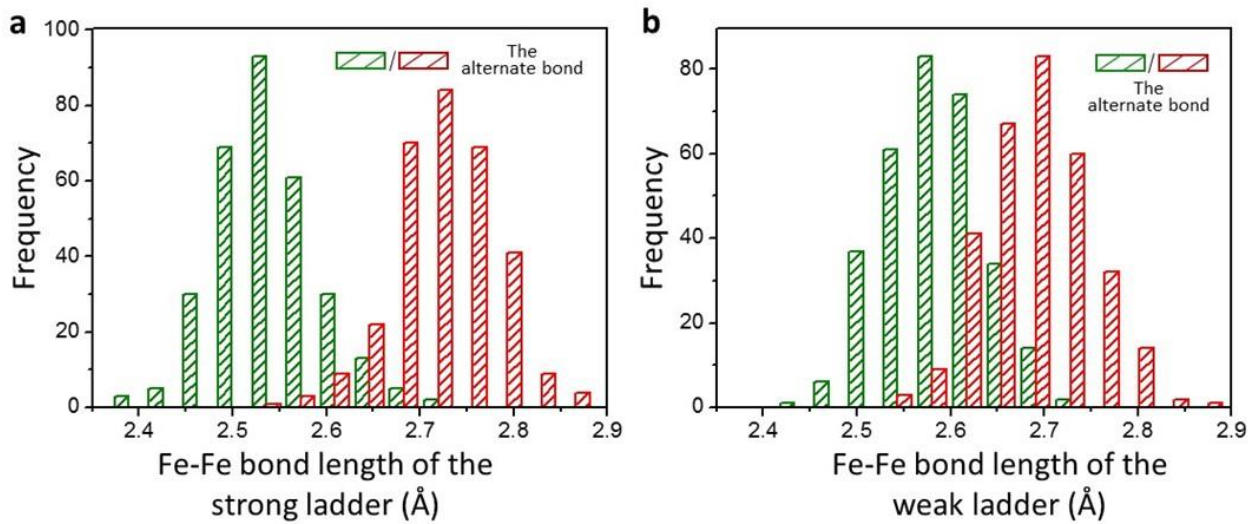

**Supplementary Figure 5** Measurement of the displacements. Based on the statistics of about 300 data, Fe-Fe bond lengths of the (a) strong ladder and (b) weak ladder are shown. According to the most probable values of distributions, the alternate Fe-Fe bond lengths in strong ladders are 2.55±0.04 Å and 2.71±0.04 Å respectively, in which the displacement of Fe atoms are stronger than that in the weak ladders where the Fe-Fe bond lengths reach 2.69±0.04 Å and 2.57±0.03 Å respectively. Strickly speaking, there are eight kinds of the Fe-Fe distance in the leg direction, since in the *Pmn2$_1$* space group the Fe-Fe rectangles are already slightly distorted to be trapezoidal. However, these trapezoidal distortion is very weak (~0.013 Å according to Ref. [27]), while the tetramerization distortion is in the range of 0.05-0.25 Å according to our STEM data. Due to the resolution limit of STEM, these eight kinds of the Fe-Fe distance degenerate as four kinds, by neglecting the tiny trapezoidal distortion.

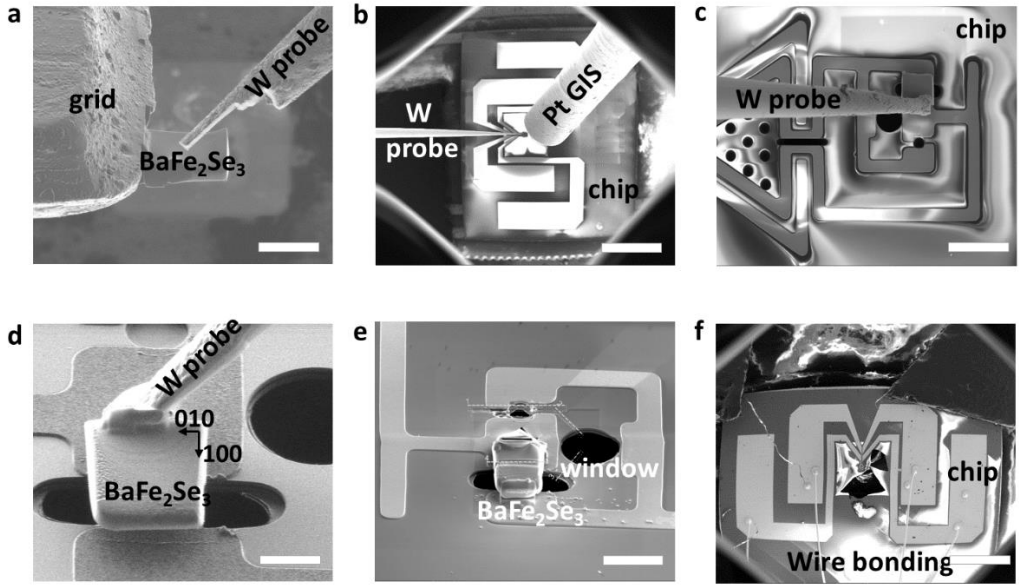

**Supplementary Figure 6** The SEM images showing the technique using a probe in order to put the sample onto a chip for testing the macroscopic properties tests. Some tests including the ferroelectric measurements along the *a*-axis and its behavior with temperature are done on these micro devices. (a) Welding the sample to tungsten probe (marked as W probe) and cutting the sample from the grid. Scale bar, 10 μm. (b) Transfer of the sample to the four-probe electricity chip and welding it onto the chip by Pt GIS. Scale bar, 1 mm. (c) Enlarged view of the end region of the chip. We place the chip on the window between electrodes which act as heat conductor or electric conductor. Scale bar, 15 μm. (d) Magnified image of the end region. Scale bar, 3 μm. (e) Polishing the surrounding area of the lamella by a Ga ions beam. Scale bar, 10 μm. (f) Connecting useful electrodes to external equipment by wire-bonding. Scale bar, 0.3 mm.

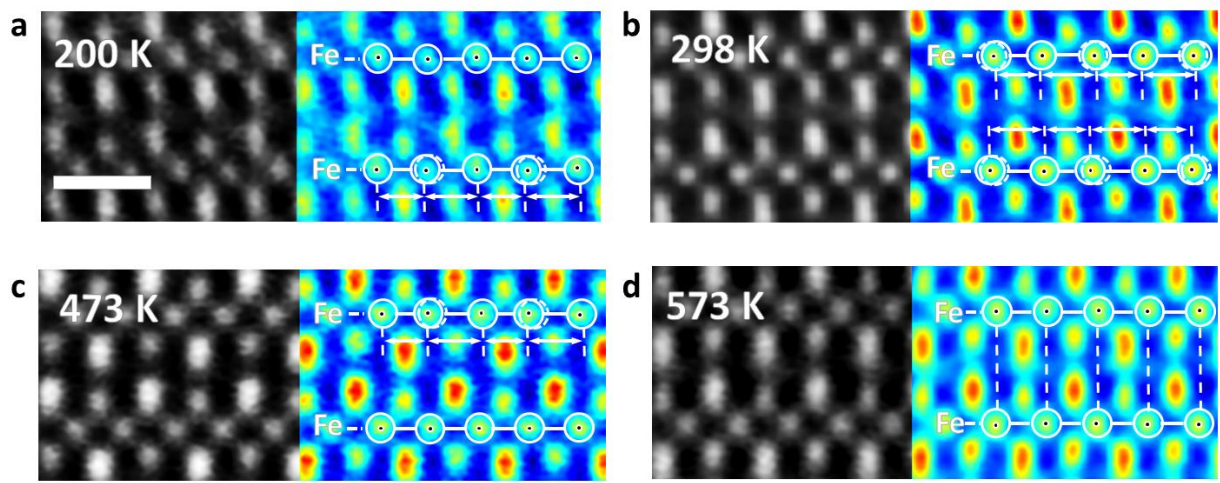

**Supplementary Figure 7** Color-enhanced HAADF-STEM images acquired at different temperatures along the *c*-axis. The strong ladder shows distinct tetramerization at (a) 200 K, (b) 298 K, and (c) 473 K, and the weak ladder reveals tetramerization at (b) 298 K. The difference of Fe-Fe bond length at high temperature (d) 573 K vanishes in both ladders indicating that in-ladder tetramerization almost drops to zero. Scale bar, 5 Å.

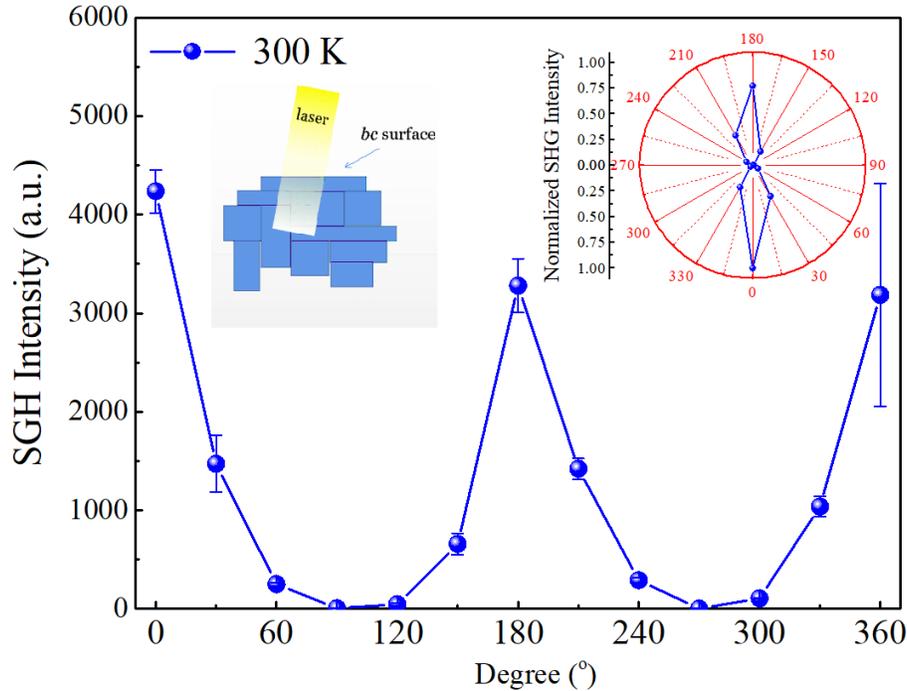

**Supplementary Figure 8** In-plane rotation of the polarization direction of the incident laser. The initial polarization (i.e. electric field component $E$) of the incident laser is along the $c$-axis. Insert (right): a polar diagram. A single axis behavior, i.e. 180º as a period, is evidenced. We repeated such rotation in several temperatures, and the 180º symmetry did not change. When $E \| b$, the signal is almost zero, which can exclude the surface contribution. Although our crystal only own large cleavage surface of $bc$ plane [see Fig. S10b], the laser of 800 nm wavelength used in SHG measurement can penetrate into the crystals partially and reflects some 'inner' information beyond the surface, as sketched in insert (left). In fact, the SHG measurements can even create small burning holes by light spots. Then the $P_a$ contribution from inner twin crystals are very possible to be detected by SHG. Considering the much larger amplitude of $P_a$ than $P_c$ (in most cases $P_a \sim$ 10-100$P_c$), a little concentration of $P_a$ domain can dominant the SHG signal, over the $P_c$'s signal. Another source of $P_a$'s signal is the canting angle of incident laser. If the incident angle is not exactly perpendicular to the $bc$ surface, the electric field component along the $a$-axis will be available.

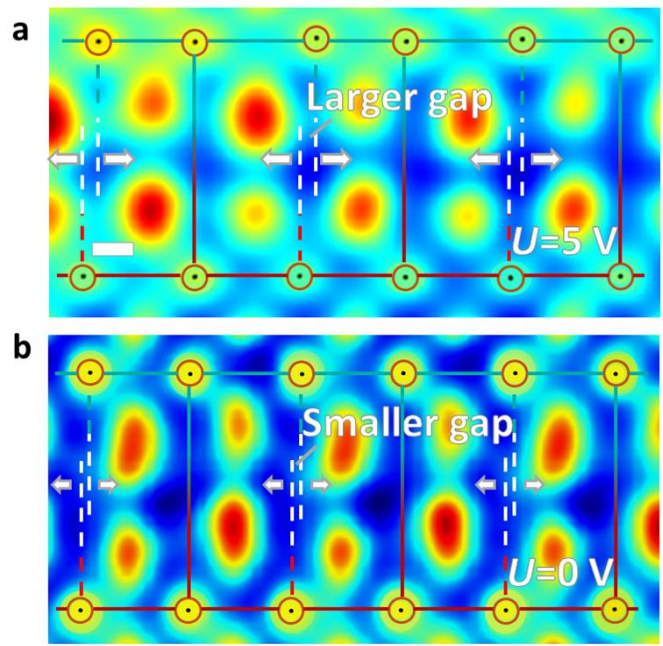

**Supplementary Figure 9** Color-enhanced HAADF-STEM images of the difference of Fe atoms displacement in the in-situ bias experiment. The neighboring Fe-Fe bond length in the weak chain under bias is equal while an enhanced tetramerization of the Fe-block is observed on the strong chain. The difference between weak and strong chain evolution is much smaller without bias. The enhanced tetramerization here indicates a corresponding enhanced polarization along the *a*-axis. Scale bar, 1 Å.

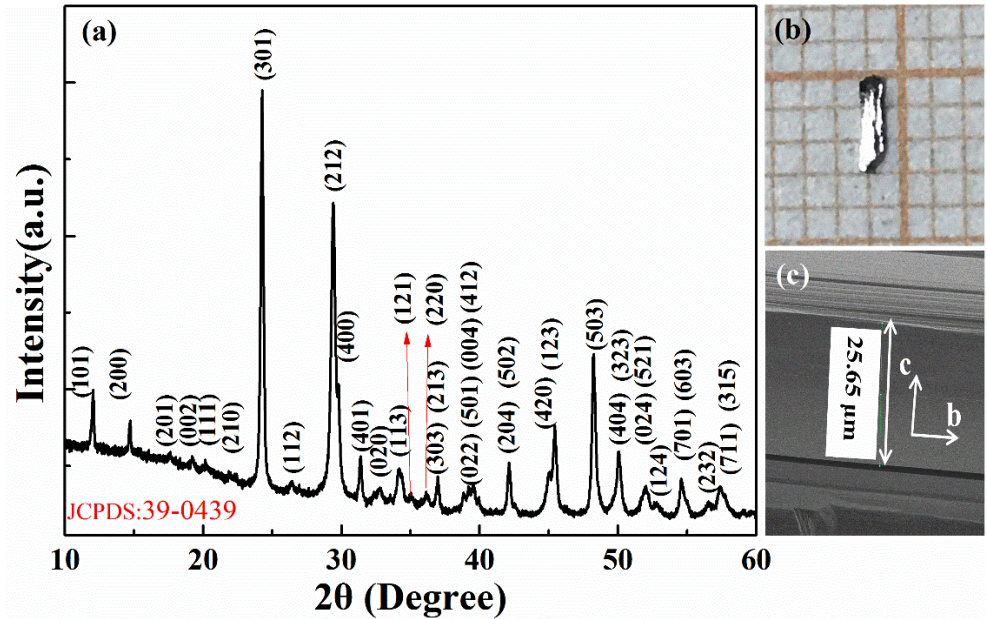

**Supplementary Figure 10** (a) X-ray powder diffraction of BaFe$_2$Se$_3$. Optical (b) and SEM (c) microscopy on BaFe$_2$Se$_3$ fiber show that the extension direction is actually along *b*-axis, thus the direction of Fe chains. Each piece of our sample (of mm size) is consisted of many loosely-contacted tiny needle-like crystals. For these crystals, the needle direction is along the *b*-axis, while the typical scale of crystals along *a*- or *c*-axis is only of μm size. Only relative large (still very small) *bc* cleavage surface is available, and it does not imply a whole single crystalline below the surface. Twin crystals are very possible below the cleavage surface (see insert of Fig. S8), considering the edge with chipped steps besides the very small cleavage surface (see the SEM image).

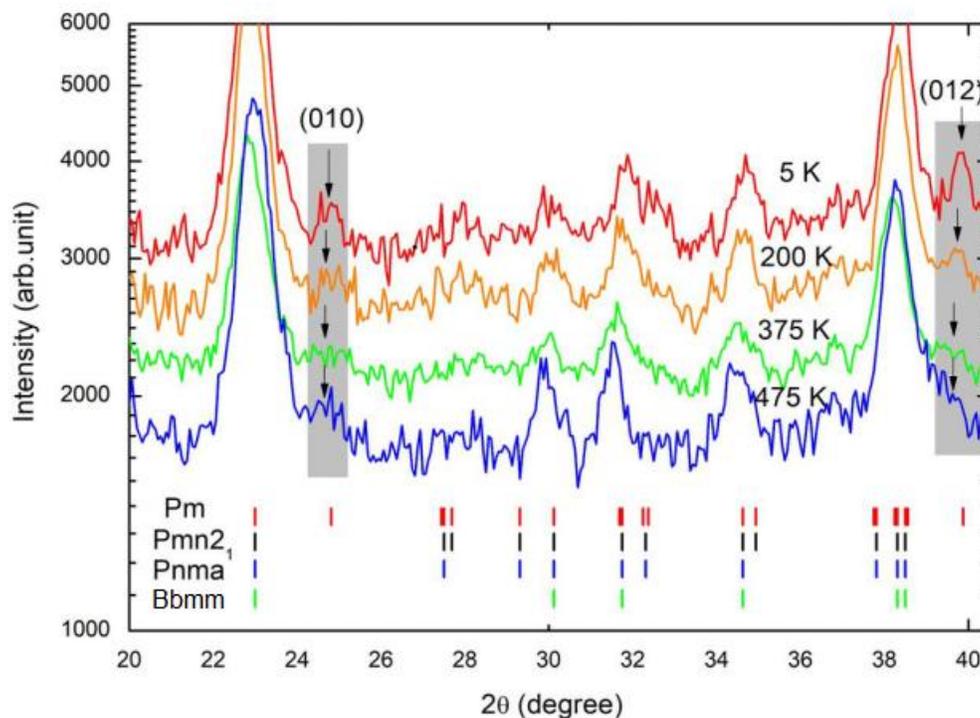

**Supplementary Figure 11** Magnified view of neutron powder diffraction of BaFe$_2$Se$_3$. The indexes of possible space groups (No. 63 *Bbmm*, No. 62 *Pnma*, No. 31 *Pmn2$_1$*, and No. 6 *Pm*) are marked. Our result supports the *Pm* space group from 475 K to 5 K. The two characteristic peaks of *Pm* space group are highlighted, which are forbidden in the *Pmn2$_1$* and *Pnma* space groups. Noting that our neutron diffraction pattern are generally in agreement with previous measurements [15, 17]. The two weak peaks which are allowed in the low symmetry space group but forbidden in the high symmetry one are close to the precision limit of our neutron spectrometer. In fact, the powder neutron diffraction data can be well refined with either the *Pm*, *Pmn2$_1$* or *Pnma* space group, as done in previous works. In short, for this material, it's hard to distinguish these space groups with neutron data only, but the neutron data can be a supplementary support to our STEM data.

132 **Supplementary Table 1** Statistics of the deviation of Fe-Fe bond length in the strong and weak
133 chain at different temperatures. The standard deviation and standard error are colored by green
134 and yellow, respectively.

| $T$ (K) | Δ Strong Chain (Å) | Δ Weak Chain (Å) |
|---|---|---|
| 200 | 0.18±0.06(0.004) | 0.03±0.08(0.005) |
| 240 | 0.15±0.05(0.003) | 0.07±0.05(0.003) |
| 270 | 0.15±0.03(0.002) | 0.08±0.04(0.002) |
| 298 | 0.16±0.04(0.002) | 0.12±0.04(0.002) |
| 343 | 0.19±0.05(0.003) | 0.16±0.06(0.004) |
| 373 | 0.20±0.04(0.002) | 0.18±0.05(0.003) |
| 423 | 0.24±0.03(0.002) | 0.07±0.03(0.002) |
| 473 | 0.17±0.05(0.002) | 0.01±0.05(0.003) |
| 573 | 0.01±0.04(0.002) | 0.01±0.05(0.003) |

Deviation of Bond length ± standard deviations (standard errors)

135